\documentclass[prl,superscriptaddress,showpacs,twocolumn]{revtex4}
\usepackage{graphicx}
\usepackage{amsmath}

\newcommand{\K}{~\mbox{K}}
\def\vo2{VO$_2$}
\def\vcro2{V$_{1-x}$Cr$_x$O$_2$}
\def\etal{{\it et al.~}}
\def\a1g{$a_{1g}$}
\def\t2g{$t_{2g}$}
\def\dparr{$d_{\parallel}$}
\def\hlda{H^{\rm{LDA}}}
\def\vk{{\bf k}}

\begin{document}

\author{ S. Biermann}
\affiliation{Centre de Physique Th{\'e}orique, Ecole Polytechnique,
91128 Palaiseau Cedex, France}
\affiliation{LPS, Universit{\'e} Paris Sud, B{\^a}t. 510, 91405 Orsay, France}
\author{ A. Poteryaev}
\affiliation{University of Nijmegen, NL-6525 ED Nijmegen, The Netherlands}
\affiliation{Centre de Physique Th{\'e}orique, Ecole Polytechnique,
91128 Palaiseau Cedex, France}
\author{ A. I. Lichtenstein}
\affiliation{Institut f{\"u}r Theoretische Physik,
Universit{\"a}t Hamburg, Jungiusstrasse 9, 20355 Hamburg, Germany}
\author{ A. Georges}
\affiliation{Centre de Physique Th{\'e}orique, Ecole Polytechnique,
91128 Palaiseau Cedex, France}
\affiliation{LPS, Universit{\'e} Paris Sud, B{\^a}t. 510, 91405 Orsay, France}

\title{Dynamical singlets and
correlation-assisted Peierls transition in VO$_2$}

\begin{abstract}

A theory of the metal-insulator transition in
vanadium dioxide from the high- temperature rutile to the low-
temperature monoclinic phase is proposed on the basis of cluster
dynamical mean field theory, in conjunction with the density
functional scheme. The interplay of strong electronic Coulomb
interactions and structural distortions, in particular the
dimerization of vanadium atoms in the low temperature phase, plays
a crucial role. We find that VO$_2$ is {\it not} a conventional Mott
insulator, but that the formation of dynamical V-V singlet pairs due to
strong Coulomb correlations is
necessary to trigger the opening of a Peierls gap.
\end{abstract}

\pacs{71.27.+a, 71.30.+h, 71.15.Ap}
\maketitle

Vanadium dioxide (\vo2) undergoes a first-order transition 
from a high-temperature
metallic phase to a low-temperature insulating phase~\cite{morin_mit_prl_1959}
at almost room-temperature ($T=340\K$). The resistivity jumps
by several orders of magnitude through this transition,
and the crystal structure changes from rutile (R-phase) at
high-temperature to monoclinic (so-called M$_1$-phase) at
low-temperature. The latter is characterized by a dimerization of
the vanadium atoms into pairs, as well as a tilting of these
pairs with respect to the c-axis.

Whether these structural changes are solely responsible for the
insulating nature of the low-T phase, or whether correlation effects also
play a role, has been a subject of much debate.
The strong dimerization, as well as the fact that the insulating
phase is non-magnetic suggests that \vo2 might be a typical case
of a Peierls insulator.
However, pioneering experimental work by Pouget \etal demonstrated that
minute amounts of Cr-substitutions~\cite{pouget_vcro2_prb_1974},
as well as, remarkably,
uniaxial stress applied to {\it pure} \vo2~\cite{pouget_uniaxial_prl_1975}
leads to a new phase (M$_2$)
in which only half of the V-atoms dimerize, while the other half
forms chains of equally-spaced atoms behaving as spin-$1/2$
Heisenberg chains. That this phase is also insulating strongly
suggests that the physics of \vo2 is very close to that of
a Mott-Hubbard insulator. Zylbersztejn and
Mott~\cite{zylbersztejn_vo2_prb_1975}, Sommers and
Doniach~\cite{sommers_vo2_ssc_1978} and Rice
\etal\cite{rice_vo2_prl_comment_1994} suggested that
Coulomb repulsion indeed plays a major role in opening the
insulating gap.

The main qualitative aspects of the electronic structure of
\vo2 have been explained long ago by
Goodenough~\cite{goodenough_vo2_jssc_1971}.
In the rutile structure
(space group $P4_{2}/mnm$), the V atoms are surrounded by
O octahedra forming an edge-sharing chain along the c-axis.
The $d$-levels of the V-ions are split
into lower lying $t_{2g}$ states and $e^{\sigma}_g$ states. The latter lie
higher in energy and are therefore empty. The tetragonal crystal
field further splits the $t_{2g}$ multiplet into an $a_{1g}$
state and an $e^{\pi}_g$ doublet ($d_{\parallel}$ and $\pi^*$
states, respectively in the terminology of
Ref.~\cite{goodenough_vo2_jssc_1971}).
The $a_{1g}$ orbitals are directed along
the c-axis, with good $\sigma$-bonding of the V-V pair along this
direction. In the monoclinic phase (space group $P2_1/c$),
the dimerization and tilting of the V-V pairs results in two
important effects. First, the \a1g (\dparr) band is split into a
lower-energy bonding combination and a higher-energy antibonding
one. Second, the V$_d$-O$_p$ antibonding $e_g^\pi$
($\pi^*$) states are pushed higher in energy, due to the tilting
of the pairs which increases the overlap of these states with
O states.
In Goodenough's picture, the single d-electron occupies
the \dparr-bonding combination, resulting in a (Peierls-like)
band gap.

Electronic structure calculations based on density functional
theory within the local density approximation (DFT-LDA)
have since provided support for this qualitative
description in terms of molecular orbitals (see the
recent work of
Eyert~\cite{eyert_vo2_annphys_2002} for an extensive
discussion and references).
Molecular dynamics calculations by
Wentzcovich \etal\cite{wentzcovich_vo2_prl_1994} with
variable cell shape successfully found the M$_1$ structure
to have the lowest total energy, with structural parameters
in reasonable agreement with experiment.
DFT-LDA calculations fail however to yield the
opening of the band gap: the top of the bonding \dparr -band is found to
overlap slightly with the bottom of the $\pi^*$-band
(only for a hypothetical structure with
larger dimerizations would the band gap fully open).
Not surprisingly, recent DFT-LDA calculations of the M$_2$
phase~\cite{eyert_vo2_annphys_2002}
also fail in producing an insulator.

This discussion makes clear that only a theoretical
treatment in which structural aspects as well as correlations
within V-V pairs are taken into account on equal footing, is able
to successfully decide on the underlying mechanism for the
metal-insulator transition in \vo2. In this Letter, we fulfill
this goal by using a cluster extension of dynamical mean-field
theory (C-DMFT)
\cite{biroli_cluster}
in combination with state-of-the-art DFT-LDA calculations within
the recently developed Nth-order muffin-tin orbital (NMTO)
\cite{nmto} implementation. This allows for a consistent
description of both the metallic rutile phase and the insulating
monoclinic phase. We find that the insulating state can be viewed
as a molecular solid of singlet dimers in the Heitler-London
(correlated) limit. While a description in terms of renormalised
Peierls bands is possible at low-energy, broad Hubbard bands
are present at higher energy.
Recent photoemission experiments can be successfully interpreted
on the basis of our results, which also yield specific predictions
for inverse-photoemission spectra.

Previous theoretical work on \vo2 based on DMFT has recently
appeared~\cite{laad_vo2,liebsch_vo2}.
These works, however, are based on a single-site DMFT
approach. This is appropriate in the metallic phase, but
not in the insulating phase. While we do find that by increasing
$U$ to unphysically high values, a Mott insulator can be induced
in a single-site DMFT approach, the formation of singlet pairs
resulting from the strong dimerization can only be captured in a
cluster extension of DMFT in which the dimers are taken as
the key unit. Only then can a non-magnetic insulator with a
spin-gap be obtained, in agreement with experiments.
C-DMFT allows for a consistent extension to the solid-state of
the simple Heitler-London picture of an isolated molecule.
Electrons can be shared between all singlets through
the self-consistent electronic bath,
resulting in a ``dynamical singlets'' description.

\begin{figure}[th]
\begin{center}
  \includegraphics[width=0.45\textwidth]{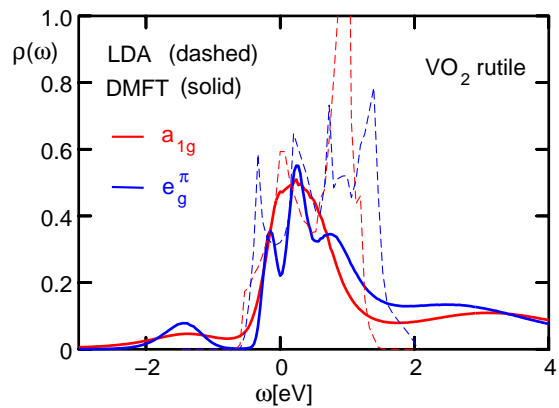}
  \end{center}
\caption{Spectral function for the rutile phase as calculated
within DMFT with $U=4$eV, $J=0.68$eV (solid lines) in comparison
to the LDA DOS (dashed lines). The red (blue) lines show the
partial contributions of the a$_{1g}$ (e$_g^{\pi}$) bands. }
\label{R_spectra}
\end{figure}

\begin{figure}[th]
\begin{center}
\includegraphics[width=0.45\textwidth]{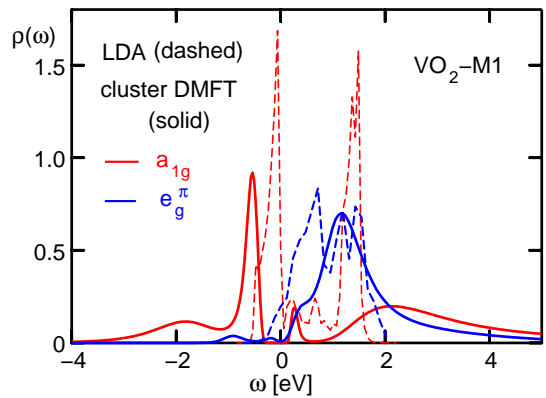}
\end{center}
\caption{Spectral function for the monoclinic phase as calculated
within cluster-DMFT with $U=4$eV, $J=0.68$eV (solid lines) in
comparison to the LDA DOS
(dashed lines). The red [blue] lines show the partial
contributions of the
a$_{1g}$ [e$_g^{\pi}$ (averaged)] bands. } \label{M1_2x2_spectra}
\end{figure}

Given the filling of one d-electron per vanadium, and the crystal-
field splitting separating the $t_{2g}$ and $e_g^\sigma$ states in
both phases, it is appropriate to work within a set of localized
V-centered $t_{2g}$ Wannier-orbitals, and to neglect the degrees
of freedom from all other bands. As in
Ref.~\cite{pavarini_d1oxides_prl}, our Wannier orbitals are
orthonormalized NMTOs, which have all partial waves other than V-
$d_{xy},d_{yz}$ and $d_{zx}$
downfolded. These notations refer to the {\it
local coordinate axes} ($z\parallel[110]$) attached to a given
V atom surrounding the oxygen octahedron.
DFT-LDA calculations followed by this downfolding procedure yield
a Hamiltonian matrix $\hlda_{mm'}(\vk)$ (of size $6\times 6$ in
the rutile phase, which has 2 V-atoms per unit cell, and
$12\times 12$ in the monoclinic phase, with $4$ V atoms per
unit cell). Our results for the DFT-LDA electronic structure in
both phases are in agreement with previously published
studies, and with the qualitative picture described above. The
partial densities of states for the $a_{1g}\equiv d_{xy}$ and
$e_g^{\pi}\equiv\{d_{yz},d_{xz}\}$ are presented in
Fig.~\ref{R_spectra} and Fig.~\ref{M1_2x2_spectra} for the rutile
and monoclinic phases, respectively. The total $t_{2g}$ bandwidth
is almost the same for both phases ($\sim 2.59$eV in rutile and
$\sim 2.56$eV in monoclinic). In the rutile phase, the single
d-electron is almost equally distributed between all three
orbital components within LDA ($0.36$ in \a1g and $0.32$ in each
of the $e_g^\pi$'s). From the Hamiltonian matrix in real
space, we can extract the hopping integrals between V
centers.
The largest one, $t_{xy,xy}=-0.31$eV is between \a1g orbitals
forming chains along the c-axis, but the hoppings $t_{xz,yz}$
within the chains and $t'_{yz,yz},t'_{xz,xz}$ are only twice
smaller ($0.17-0.19$eV). Given that there are $8$ next-nearest
neighbours of this type, and only $2$ along the chains, the
properties of the rutile metal are therefore fairly isotropic.
The electronic structure drastically changes in the monoclinic
phase (Fig.~\ref{M1_2x2_spectra}). The \a1g state is split into a
bonding combination below the Fermi level, and an antibonding
combination higher in energy. The splitting between the bonding
and antibonding states is set by the intra-dimer hopping, which
we find to be $t^{\rm{intra}}_{xy,xy}=-0.68$~eV, hence a
splitting of order $2 t_{xy,xy}\sim 1.4$~eV. The two \a1g peaks
in the density of states (DOS) are very narrow, corresponding to
the very small inter-dimer hopping $t^{\rm inter}_{xy,xy}\simeq
-0.03$~eV. The intra-dimer hopping is by far the dominant one in
this phase, with the second largest being $t^{\prime}_{xz,yz}$ in the
$[111]$ direction ($\sim 0.22$~eV) and all others much
smaller. Also, the $e_{g}^\pi$ states are pushed higher in energy
in the monoclinic phase, so that the LDA occupancies are now
$0.74$ for the $a_{1g}$ (bonding) band and only $0.12$ and $0.14$
for the $d_{yz}$ and $d_{xz}$, respectively. Still, these bands
overlap weakly and the LDA fails to open the insulating gap ($\sim
0.6$~eV experimentally).

We use the LDA-NMTO Hamiltonian $\hlda_{mm'}(\vk)$ as a starting point
for the construction of a multi-band Hubbard Hamiltonian of the
form of Eq.~(1) in Ref.~\cite{pavarini_d1oxides_prl}
involving direct and exchange terms of the
screened \emph{on-site} Coulomb interaction $U_{mm^{\prime }}$
and $J_{mm^{\prime }}$, with the parametrization
$U_{mm}=U,\ U_{mm^{\prime }}=U-2J$
and $J_{mm^{\prime }}=J$ for $m\mathrm{\neq }m^{\prime }$.
We assume double counting corrections to be orbital-independent
within the t$_{2g}$ manifold, thus resulting in a simple shift of
the chemical potential.
Recently, it has become feasible to solve this many-body
Hamiltonian using the dynamical mean-field approximation (DMFT)
and to obtain realistic physical properties, even when all
off-diagonal terms in orbital space in the local
self-energy $\Sigma_{mm'}$
are retained~\cite{pavarini_d1oxides_prl}. Here, we go one step
further by including also non-local terms in the self-energy. The
latter are constructed from a cluster LDA+DMFT
treatment~\cite{poteryaev_Ti2O3_prl_2004}. More precisely, instead of calculating the
self-energy from a local impurity model embedding one single atom
in a self-consistent bath, a {\it pair} of V atoms in the
complex bath is explicitly considered.
The self-energy $\Sigma_{mm'}^{i_cj_c}$ and (Weiss) dynamical
mean-field ${\cal G}_{mm'}^{i_cj_c}$ become matrices in both the
orbital indices $m,m'=(xy,yz,xz)$ and the intra-dimer site indices
$i_c,j_c=1,2$. The inter-dimer components of the self-energy as
well as long-range correlations are neglected in this C-DMFT
scheme.
Using the crystal symmetries in the monoclinic
phase, a $12\times 12$ block-diagonal
self-energy matrix is constructed from the
$6\times 6$ matrix $\Sigma_{mm'}^{i_cj_c}$ so that the C-DMFT
approximation to the self-energy takes in this case the form~:
\begin{eqnarray}\label{SigmaMatrix}
\Sigma
=
\left(
\begin{array}{cccc}
\hat{\Sigma}_{\rm 11} & \hat{\Sigma}_{\rm 12} & 0 & 0\\
\hat{\Sigma}_{\rm 21} & \hat{\Sigma}_{\rm 22} & 0 & 0 \\
0 & 0 & \hat{\Sigma}_{\rm 11} & \hat{\Sigma}_{\rm 12} \\
0 & 0 & \hat{\Sigma}_{\rm 21} & \hat{\Sigma}_{\rm 22}  \\
\end{array}
\right)
\end{eqnarray}
where $\hat{\Sigma}_{\rm 12}$ [$\hat{\Sigma}_{\rm 11}$]
denotes the 3 $\times$ 3 intersite [on-site] self-energy matrix
in the space of the \t2g orbitals. This is then
combined with $\hlda_{mm'}(\vk)$ in order to obtain the Green's
function at a given $\mathbf{k}$-point. After summation over
$\mathbf{k}$, the intra-dimer block of the Green's function
$G_{mm'}^{i_cj_c}$ is extracted and used in the C-DMFT
self-consistency condition. The 6-orbital impurity problem is
solved by a numerically exact quantum Monte Carlo (QMC) scheme
using up to 100 slices in
imaginary time at temperatures down to 770 K. 
From the Green's function on the imaginary-time axis we
calculate the spectral function by using the
maximum entropy technique.

We first discuss our results for the rutile phase
(Fig.~\ref{R_spectra}). In this phase, we found that the
results of single-site and cluster-DMFT calculations are
indistinguishable for all practical purposes. Correlations
reduce the {\it total} (occupied and unoccupied)
bandwidth corresponding to the coherent
part of the spectral density from $2.59$~eV in LDA down to
about $1.8$~eV (see Fig.~\ref{R_spectra}). 
This is consistent with the quasiparticle
weight that we extract from the self-energy: $Z\simeq 0.66$.
Hence, the metallic phase of \vo2 can be characterized as
a metal with an intermediate level of correlations.
We observe however, that the {\it occupied bandwidth}
is barely modified ($\sim 0.5 eV$ in both the LDA and
our results). This finding is in agreement with
photoemission experiments~\cite{okazaki_vo2_pes_prb_2004,
koethe_vo2},
and our work demonstrates that it is compatible
with a local self-energy.
A prominent quasi-particle coherence peak is found close
to the Fermi level. This has recently been demonstrated 
experimentally in high photon energy bulk-sensitive
photoemission experiments \cite{koethe_vo2} (and is also
suggested by low-photon energy PES provided surface
contributions are substracted \cite{okazaki_vo2_pes_prb_2004}).
Hubbard bands are apparent at higher energies, with a
rather weak lower Hubbard band about $-1.5$~eV below the
Fermi level and
a more pronounced upper Hubbard band at about $+2.5-3$~eV.
The former has been observed in photoemission
experiments~\cite{shin_pes_vanadates_prb_1990,
okazaki_vo2_pes_prb_2004, 
koethe_vo2},
while the latter constitutes a prediction for BIS experiments.
Finally, correlations result in a slight increase of the
occupancy of the $a_{1g}$ band $(0.42)$ relative to the
$e_g^\pi$ ones $(0.29,0.29)$,
in comparison to LDA: $(0.36;0.32,0.32)$.

\begin{figure}[th]
\begin{center}
\includegraphics[width=0.43\textwidth]{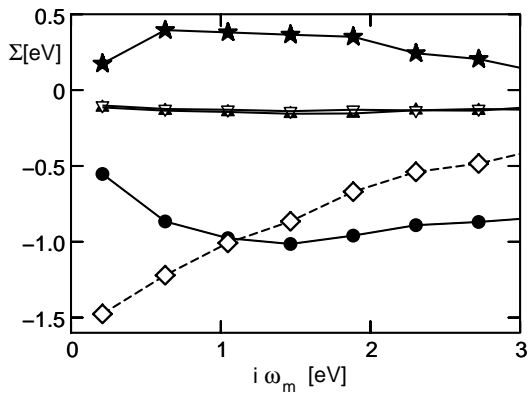}
\end{center}
\caption{Self-energies 
$\Sigma(i\omega_n)$
from the C-DMFT calculation (for U=4 eV, J=0.68 eV)
for the M$_1$ phase,
at low (imaginary)
frequency. Circles [triangles]: imaginary part of the
on-site diagonal \a1g [$e_g^\pi$] component.
Diamonds [stars]: real [imaginary] part of the
inter-site \a1g component.
}
\label{fig:self}
\end{figure}

Figure \ref{M1_2x2_spectra} displays the spectral functions for the
monoclinic phase calculated within cluster-DMFT in comparison
with the LDA DOS. The key point is that inclusion of non-local
self-energy effects succeeds in opening up a gap of about 0.6 eV,
in reasonable agreement with experiments.
Our calculations
also yield a large redistribution of the electronic
occupancies in favor of the \a1g orbital which now carries
$0.8$ electrons per vanadium (with only $\sim(0.1,0.1)$ remaining
in the $e_g^\pi$ orbitals). This charge redistribution is a
common feature of many models of \vo2, whether correlations-
or band-driven, and is also observed in experiments.
In the present context, the fact that the single electron
occupies almost entirely the \a1g orbital, together with the fact
that $U$ is larger than the intra-dimer hopping (itself much
bigger than other hoppings), means that the ground-state of each
dimer is close to the Heitler-London limit. In this limit, one
has two electrons with opposite spins on each site forming a
singlet state, rather than an uncorrelated wave-function in which
two electrons (per dimer) are placed in the bonding combination of
atomic orbitals (hence a large double occupancy). The transition
into the insulating state is facilitated by the Heitler-London
stabilisation energy. This confirms the early proposal of Sommers
and Doniach~\cite{sommers_vo2_ssc_1978}. As a result, the
dominant excitation when adding an electron {\it in the \a1g
orbital} costs an energy which is set by $U$. This is apparent on
our spectra: the antibonding combination of \a1g orbitals
corresponding to the narrow peak at $\sim +1.5$~eV present in the
LDA calculation is replaced by a broad upper Hubbard band
centered at $\sim +2.2$~eV (the precise value of course depends
on the choice of $U$). Correspondingly, there is a (weak) lower
Hubbard band at $\sim -1.8$~eV. The {\it low-energy} nature of
this singlet insulator, however, is quite different from that of
a standard Mott insulator in which local moments are formed in the
insulating state. This is manifested by the fact that the
low-frequency behaviour of the on-site component of the
self-energy associated with the \a1g orbital (Fig.~\ref{fig:self})
is linear in frequency
$\Sigma_{11}=\Sigma_{11}(0)+(1-1/Z_i)\omega+\cdots$, in contrast
to the local-moment Mott insulator in which $\Sigma_{11}\propto
1/\omega$ diverges. As a result, the spectrum displays a narrow
\a1g peak, at an energy $\sim -0.8$~eV below the gap, which
carries most of the spectral weight for $\omega<0$. This peak
should not be interpreted as an incoherent lower Hubbard band,
but rather as a quasiparticle which has been gapped out ($Z_i$
can be interpreted as the spectral weight of the gapped
low-energy quasiparticle in the insulator). Hence, at low-energy,
the physics is that of a renormalised Peierls insulator
(analogously, a correlated Kondo insulator can be viewed as a
renormalised hybridisation-gap insulator at low-energy). The
bonding-antibonding splitting is renormalized down by correlations.
Indeed, a weaker \a1g
peak is visible in our spectra at the upper gap edge, at an
energy considerably smaller than the antibonding peak in the LDA
DOS. The $e_g^\pi$ band, in contrast to the \a1g, is weakly
correlated, as evident
from the self-energy in Fig.~\ref{fig:self} and expected from the
low electron occupancy. Its bottom lies in the same energy range than
the ``renormalised'' antibonding peak, so that the gap can as 
well be considered to open between the \a1g and the $e_g^\pi$ band.

\begin{figure}[th]
\begin{center}
  \includegraphics[width=0.43\textwidth]{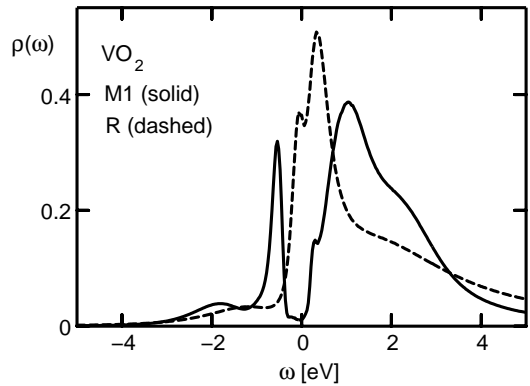}
  \end{center}
\caption{Spectral function of the t$_{2g}$ orbitals
for rutile and monoclinic phases as calculated within DMFT/cluster-DMFT
with $U=4$eV, $J=0.68$eV. The Hubbard band in the rutile phase is
located at about $-1.2$ eV, whereas the pronounced peak at $-0.8$eV
in the monoclinic phase 
corresponds to the gapped quasi-particle (see text).
}
\label{fig:total_dos}
\end{figure}

Our findings for the spectral function explain the recent
photoemission data of T. Koethe \etal \cite{koethe_vo2}. 
These
authors made the puzzling observation that a prominent peak below
the gap appears in the insulator but at an energy ($\sim
-0.8$~eV) which is shifted {\it towards positive energy} in comparison
to the (weak) lower Hubbard band measured in the metallic phase
($\sim -1.2$~eV). We propose that this peak is actually the
coherent quasiparticle peak at the bottom of the gap, rather than
the lower Hubbard band. The latter is present at higher
(negative) energy but remains rather weak even in the insulator.
Fig.~\ref{fig:total_dos} compares the total $t_{2g}$ spectral
functions calculated for the metallic and the insulating phases.

Finally, let us mention that we have discussed the
results obtained for a specific choice of the interaction
parameters $U=4$~eV and $J=0.68$~eV. We have actually performed
calculations for other choices. In particular, we found that it
is possible to stabilize an insulating state within C-DMFT for
smaller values of $U$ (e.g $U=2$~eV), but only if the Hund's
coupling $J$ is taken to be small (in which case it is easier
to redistribute the charge towards the \a1g orbital). Also, we
emphasize that a {\it single-site} DMFT calculation can lead to
an insulating state for rather large values of $U$ ($U\geq5$~eV).
However, this insulating solution is a conventional Mott
insulator with a local moment, and therefore would display a
large magnetic susceptibility which does not correspond to the
actual physics of insulating \vo2.

In conclusion, we have presented LDA+CDMFT calculations for
vanadium dioxide. Both the metallic rutile phase and the
insulating monoclinic phase are correctly captured by this
approach.

\acknowledgements
We are grateful to J-P. Pouget, C. Noguera, F. Finocchi,
V. Eyert, A. Fujimori and L.H. Tjeng
for useful discussions, and to the KITP Santa Barbara
for hospitality and support (NSF Grant PHY99-07949).
S.B and A.G acknowledge funding from CNRS and Ecole Polytechnique,
and A.P from FOM.
Computing time was provided by IDRIS (Orsay) under project No. 041393.

\bibliographystyle{apsrev}

\begin{thebibliography}{19}
\expandafter\ifx\csname natexlab\endcsname\relax\def\natexlab#1{#1}\fi
\expandafter\ifx\csname bibnamefont\endcsname\relax
  \def\bibnamefont#1{#1}\fi
\expandafter\ifx\csname bibfnamefont\endcsname\relax
  \def\bibfnamefont#1{#1}\fi
\expandafter\ifx\csname citenamefont\endcsname\relax
  \def\citenamefont#1{#1}\fi
\expandafter\ifx\csname url\endcsname\relax
  \def\url#1{\texttt{#1}}\fi
\expandafter\ifx\csname urlprefix\endcsname\relax\def\urlprefix{URL }\fi
\providecommand{\bibinfo}[2]{#2}
\providecommand{\eprint}[2][]{\url{#2}}
 
\bibitem[{\citenamefont{Morin}(1959)}]{morin_mit_prl_1959}
\bibinfo{author}{\bibfnamefont{F.~J.} \bibnamefont{Morin}},
  \bibinfo{journal}{Phys. Rev. Lett.} \textbf{\bibinfo{volume}{3}},
  \bibinfo{pages}{34} (\bibinfo{year}{1959}).
 
\bibitem[{\citenamefont{{Pouget} et~al.}(1974)\citenamefont{{Pouget},
  {Launois}, {Rice}, {Dernier}, {Gossard}, {Villeneuve}, and
  {Hagenmuller}}}]{pouget_vcro2_prb_1974}
\bibinfo{author}{\bibfnamefont{J.~P.} \bibnamefont{{Pouget}}},
  \bibnamefont{et al.}
  \bibinfo{journal}{Phys. Rev. B} \textbf{\bibinfo{volume}{10}},
  \bibinfo{pages}{1801} (\bibinfo{year}{1974}).
 
\bibitem[{\citenamefont{{Pouget} et~al.}(1975)\citenamefont{{Pouget},
  {Launois}, {D'Haenens}, {Merenda}, and {Rice}}}]{pouget_uniaxial_prl_1975}
\bibinfo{author}{\bibfnamefont{J.~P.} \bibnamefont{{Pouget}}},
  \bibnamefont{et al.}
  \bibinfo{journal}{Phys. Rev. Lett.}
  \textbf{\bibinfo{volume}{35}}, \bibinfo{pages}{873} (\bibinfo{year}{1975}).
 
\bibitem[{\citenamefont{Zylbersztejn and
  Mott}(1975)}]{zylbersztejn_vo2_prb_1975}
\bibinfo{author}{\bibfnamefont{A.}~\bibnamefont{Zylbersztejn}}
  \bibnamefont{and} \bibinfo{author}{\bibfnamefont{N.}~\bibnamefont{Mott}},
  \bibinfo{journal}{Phys. Rev. B} \textbf{\bibinfo{volume}{11}}
  (\bibinfo{year}{1975}).
 
\bibitem[{\citenamefont{Sommers and Doniach}(1978)}]{sommers_vo2_ssc_1978}
\bibinfo{author}{\bibfnamefont{C.}~\bibnamefont{Sommers}} \bibnamefont{and}
  \bibinfo{author}{\bibfnamefont{S.}~\bibnamefont{Doniach}},
  \bibinfo{journal}{Solid State Commun.} \textbf{\bibinfo{volume}{28}},
  \bibinfo{pages}{133} (\bibinfo{year}{1978}).
 
\bibitem[{\citenamefont{Rice et~al.}(1994)\citenamefont{Rice, Launois, and
  Pouget}}]{rice_vo2_prl_comment_1994}
\bibinfo{author}{\bibfnamefont{T.~M.} \bibnamefont{Rice}},
  \bibinfo{author}{\bibfnamefont{H.}~\bibnamefont{Launois}}, \bibnamefont{and}
  \bibinfo{author}{\bibfnamefont{J.~P.} \bibnamefont{Pouget}},
  \bibinfo{journal}{Phys. Rev. Lett.} \textbf{\bibinfo{volume}{73}},
  \bibinfo{pages}{3042} (\bibinfo{year}{1994}).
 
\bibitem[{\citenamefont{Goodenough}(1971)}]{goodenough_vo2_jssc_1971}
\bibinfo{author}{\bibfnamefont{J.}~\bibnamefont{Goodenough}},
  \bibinfo{journal}{J. Solid State Chem.} \textbf{\bibinfo{volume}{3}}
  (\bibinfo{year}{1971}).
 
\bibitem[{\citenamefont{Eyert}(2002)}]{eyert_vo2_annphys_2002}
\bibinfo{author}{\bibfnamefont{V.}~\bibnamefont{Eyert}}, \bibinfo{journal}{Ann.
  Phys. (Leipzig)} \textbf{\bibinfo{volume}{11}}, \bibinfo{pages}{9}
  (\bibinfo{year}{2002}).
 
\bibitem[{\citenamefont{Wentzcovitch et~al.}(1994)\citenamefont{Wentzcovitch,
  Schulz, and Allen}}]{wentzcovich_vo2_prl_1994}
\bibinfo{author}{\bibfnamefont{R.~M.} \bibnamefont{Wentzcovitch}},
  \bibinfo{author}{\bibfnamefont{W.}~\bibnamefont{Schulz}}, \bibnamefont{and}
  \bibinfo{author}{\bibfnamefont{P.}~\bibnamefont{Allen}},
  \bibinfo{journal}{Phys. Rev. Lett.} \textbf{\bibinfo{volume}{72}},
  \bibinfo{pages}{3389}
  (\bibinfo{year}{1994}).
 
\bibitem[{\citenamefont{Biroli et~al.}(2003)\citenamefont{Biroli, Parcollet,
  and Kotliar}}]{biroli_cluster} For recent reviews see
\bibinfo{author}{\bibfnamefont{G.}~\bibnamefont{Biroli}},
  \bibinfo{author}{\bibfnamefont{O.}~\bibnamefont{Parcollet}},
  \bibnamefont{and} \bibinfo{author}{\bibfnamefont{G.}~\bibnamefont{Kotliar}}
  (\bibinfo{year}{2003}), \bibinfo{note}{cond-mat/0307587~;}
%
\bibinfo{author}{\bibfnamefont{T.}~\bibnamefont{Maier}},
  \bibinfo{author}{\bibfnamefont{M.}~\bibnamefont{Jarrell}},
  \bibinfo{author}{\bibfnamefont{T.}~\bibnamefont{Pruschke}}, \bibnamefont{and}
  \bibinfo{author}{\bibfnamefont{M.}~\bibnamefont{Hettler}}
  (\bibinfo{year}{2004}), \bibinfo{note}{cond-mat/0404055~;}
%
\bibinfo{author}{\bibfnamefont{A.}~\bibnamefont{Lichtenstein}},
  \bibinfo{author}{\bibfnamefont{M.}~\bibnamefont{Katsnelson}}, \bibnamefont{and}
  \bibinfo{author}{\bibfnamefont{G.}~\bibnamefont{Kotliar}}
  (\bibinfo{year}{2002}), \bibinfo{note}{cond-mat/0211076}
.
 
\bibitem[{\citenamefont{Andersen and Saha-Dasgupta}(2000)}]{nmto}
\bibinfo{author}{\bibfnamefont{O.}~\bibnamefont{Andersen}} \bibnamefont{and}
  \bibinfo{author}{\bibfnamefont{T.}~\bibnamefont{Saha-Dasgupta}},
  \bibinfo{journal}{Phys. Rev. B} \textbf{\bibinfo{volume}{62}},
  \bibinfo{pages}{16219} (\bibinfo{year}{2000}).
 
\bibitem[{\citenamefont{Laad et~al.}(2004)\citenamefont{Laad, Craco, and
  M{\"u}ller-Hartmann}}]{laad_vo2}
\bibinfo{author}{\bibfnamefont{M.}~\bibnamefont{Laad}},
  \bibinfo{author}{\bibfnamefont{L.}~\bibnamefont{Craco}}, \bibnamefont{and}
  \bibinfo{author}{\bibfnamefont{E.}~\bibnamefont{M{\"u}ller-Hartmann}},
  \bibinfo{journal}{cond-mat/030486 (2003) and 0409027}
  (\bibinfo{year}{2004}).
                                                                                
\bibitem[{\citenamefont{Liebsch and Ishida}(2003)}]{liebsch_vo2}
\bibinfo{author}{\bibfnamefont{A.}~\bibnamefont{Liebsch}} \bibnamefont{and}
  \bibinfo{author}{\bibfnamefont{H.}~\bibnamefont{Ishida}},
  \bibinfo{journal}{cond-mat/0310216}  (\bibinfo{year}{2003}).
 
\bibitem[{\citenamefont{Pavarini et~al.}(2004)\citenamefont{Pavarini, Biermann,
  Poteryaev, Lichtenstein, Georges, and Andersen}}]{pavarini_d1oxides_prl}
\bibinfo{author}{\bibfnamefont{E.}~\bibnamefont{Pavarini}},
\bibnamefont{et al.}
  \bibinfo{journal}{Phys. Rev. Lett.} \textbf{\bibinfo{volume}{92}},
  \bibinfo{pages}{176403} (\bibinfo{year}{2004}).
 
\bibitem[{\citenamefont{Poteryaev et~al.}(2004)\citenamefont{Poteryaev,
  Lichtenstein, and Kotliar}}]{poteryaev_Ti2O3_prl_2004}
\bibinfo{author}{\bibfnamefont{A.}~\bibnamefont{Poteryaev}},
  \bibinfo{author}{\bibfnamefont{A.}~\bibnamefont{Lichtenstein}},
  \bibnamefont{and} \bibinfo{author}{\bibfnamefont{G.}~\bibnamefont{Kotliar}},
  \bibinfo{journal}{Phys. Rev. Lett.} \textbf{\bibinfo{volume}{93}},
  \bibinfo{pages}{086401} (\bibinfo{year}{2004}).
 
\bibitem[{\citenamefont{{Okazaki} et~al.}(2004)\citenamefont{{Okazaki},
  {Wadati}, {Fujimori}, {Onoda}, {Muraoka}, and
  {Hiroi}}}]{okazaki_vo2_pes_prb_2004}
\bibinfo{author}{\bibfnamefont{K.}~\bibnamefont{{Okazaki}}},
  \bibnamefont{et al.}
  \bibinfo{journal}{Phys. Rev. B} \textbf{\bibinfo{volume}{69}},
  \bibinfo{pages}{165104} (\bibinfo{year}{2004})
\bibnamefont{;}
  \bibinfo{author}{\bibfnamefont{K.}~\bibnamefont{Okazaki}},
  \bibinfo{journal}{PhD thesis, University of Tokyo}
  (\bibinfo{year}{2002}).

 
\bibitem[{\citenamefont{Koethe et~al.}(2004)\citenamefont{Koethe, Hu,
  Sch{\"u}{\ss}ler-Langeheine, Tjernberg, Venturini, Brookes, Reichelt, and
  Tjeng}}]{koethe_vo2}
\bibinfo{author}{\bibfnamefont{T.}~\bibnamefont{Koethe}},
  \bibinfo{author}{\bibfnamefont{Z.}~\bibnamefont{Hu}},
  \bibinfo{author}{\bibfnamefont{C.}~\bibnamefont{Sch{\"u}{\ss}ler-Langeheine}%
}, \bibinfo{author}{\bibfnamefont{O.}~\bibnamefont{Tjernberg}},
  \bibinfo{author}{\bibfnamefont{F.}~\bibnamefont{Venturini}},
  \bibinfo{author}{\bibfnamefont{N.}~\bibnamefont{Brookes}},
  \bibinfo{author}{\bibfnamefont{W.}~\bibnamefont{Reichelt}}, \bibnamefont{and}
  \bibinfo{author}{\bibfnamefont{L.H.}~\bibnamefont{Tjeng}}
  (\bibinfo{year}{2004}), \bibinfo{note}{private comm. and to be published}.

\bibitem[{\citenamefont{{Shin} et~al.}(1990)\citenamefont{{Shin}, {Suga},
  {Taniguchi}, {Fujisawa}, {Kanzaki}, {Fujimori}, {Daimon}, {Ueda}, {Kosuge},
  and {Kachi}}}]{shin_pes_vanadates_prb_1990}
\bibinfo{author}{\bibfnamefont{S.}~\bibnamefont{{Shin}}},
  \bibnamefont{et al.}
  \bibinfo{journal}{Phys. Rev. B} \textbf{\bibinfo{volume}{41}},
  \bibinfo{pages}{4993} (\bibinfo{year}{1990}).

 
\end{thebibliography}

\end{document}